\input phyzzx\def\e{\adveq\eqno{\rm (\chapterlabel.\the\equanumber)}}

\def\adveq{\global\advance\equanumber by 1}
\def\myeq{{\rm \chapterlabel.\the\equanumber}}

\def\semidirect{\mathrel{\raise0.04cm\hbox{${\scriptscriptstyle |\!}$
\hskip-0.175cm}\times}}

\def\mod{\mathop{\rm mod}\nolimits}

\def\ref#1{$^{[#1]}$}

\def\r#1{$[\rm#1]$}

\def\e{\adveq\eqno{\rm (\chapterlabel.\the\equanumber)}}

\def\adveq{\global\advance\equanumber by 1}
\def\myeq{{\rm \chapterlabel.\the\equanumber}}

\def\semidirect{\mathrel{\raise0.04cm\hbox{${\scriptscriptstyle |\!}$
\hskip-0.175cm}\times}}

\def\mod{\mathop{\rm mod}\nolimits}

\def\ref#1{$^{[#1]}$}

\def\r#1{$[\rm#1]$}

\def\half{{1\over2}}

\def\e{\adveq\eqno{\rm (\the\equanumber)}}

\date{April, 2022}
\date{April, 2022}
\titlepage
\title{Generalized Rogers Ramanujan Expressions for Some Non--Singlet Twisted Affine Algebras.}
\author{Doron Gepner}
\vskip20pt
\line{\it\hfill  Department of Particle Physics and Astrophysics, Weizmann Institute, \hfill}
\line{\it\hfill Rehovot 76100,  Israel\hfill} 

\abstract
Hatayama et al. described generalized Rogers--Ramanujan (GRR) expressions for the 
string functions of the 
singlet representation of twisted affine algebras. We give here such GRR expressions for some
non-singlet string functions. In the case of the algebra $A_2^{(2)}$ this gives all the
string functions. We verify these expressions using Freudenthal--Kac formula.
\endpage 

Generalized Rogers--Ramanujan (GRR) identities are of great mathematical interest. In physics
they appeared as expressions for the characters of two--dimensional conformal field theories
\REF\Bel{A.A. Belavin, A.M. Polyakov and A.B. Zamolodchikov, Nucl. Phys. B 241 (1984) 333.}
\r\Bel. Of special interest are the conformal field theories associated to Lie groups
called generalized parafermions
\REF\Gep{D. Gepner, Nucl. Phys. `B 290 (1987) 10.}\r\Gep. The characters of these theories 
were shown there to be given by the string functions of affine Lie algebras.
These are generating functions of a string of weights in some representation
\REF\Kac{V.G. Kac, ``Infinite dimensional Lie algebras", Cambridge University Press
(1990)} \r\Kac. Thus, we seek GRR identities for the string functions of affine algebras.
In a seminal work, Lepowsky and Primc found such expressions for $SU(2)$
\REF\Lep{J. Lepowsky and M. Primc,  Contemporary Mathematics, Vol. 46 AMS (1985).}
\r\Lep. Later, GRR expressions for the untwisted affine algebras were given for the singlet representation
\REF\Kun{A. Kuniba, T. Nakanishi and J. Suzuki, Mod.Phys.Lett. A 8 (1993) 1649.}
\r\Kun, and generalized to some other representations in
\REF\Geptwo{D. Gepner, Lett. Math. Phys. 105 (2015) 6, 769.}\r\Geptwo. These GRR expressions,
for the untwisted algebras, were recently proved in
\REF\Pri{M. Butorac, S. Kozic and M. Primc, Math. Zeit. 298 (2021) 1003.}\r\Pri.
The GRR for the twisted affine algebras and the singlet representation were given
by Hatayama et al. 
\REF\Hata{G. Hatayama, A. Kuniba, M. Okado, T. Takagi and Z. Tsuboi, Prog. Math. Phys.
23 (2002) 205.}\r\Hata.
These GRR expressions were just proved for all the twisted algebras except $A_{2l}^{(2)}$
\REF\Okado{M. Okado and R. Takenaka, arXiv: 2109.08892 (2021).}\r\Okado, and 
for the singlet representation.
Our purpose here is to give expressions for some other representations of the twisted affine algebras. We find such GRR
expressions for some non--singlet representations. In the case of $A_2^{(2)}$ algebra we find
all the string functions. We verify our conjecture for low grades using the computer program ALGEBRA,
which is based on Freudenthal--Kac formula \r\Kac.
In the case of singlet representations this verification was done in 
\REF\Gen{A. Genish and D. Gepner, Mod. Phys. Lett. A, Vol. 32, No. 21 (2017) 1750110.}
\r\Gen.

The twisted affine algebras were classified by Kac \r\Kac. The degree of the
twist is given by $r=2$ or $3$. The algebras are:
$$A_2^{(2)}:\,A_1,\quad A_{2l}^{(2)}:\, B_l,\quad A^{(2)}_{2l-1}:\, C_l,\quad D^{(2)}_{l+1}:\, B_l,\quad
E_6^{(2)}:\, F_4,\quad D_4^{(3)}:\, G_2,\e$$
where the superscript of the algebra is the degree of the twist $r$ and the algebra on the right is 
the algebra invariant under the twist, $G_{\bar 0}$. We denote by $\alpha_a$ the simple roots of $G_{\bar 0}$.
The scalar product matrix is normalized so that the long root square is $2 r$. The basis that we choose 
for the simple roots is as follows:
$$A_2^{(2)}:\, \alpha_1=2\epsilon_1,\e$$
$$A_{2 l}^{(2)}(l\geq 2):\, \alpha_i=\sqrt 2(\epsilon_i-\epsilon_{i+1}),\quad i=1,2,\ldots, l-1,\quad \epsilon_l=\sqrt2 \epsilon_l,\e$$

$$A^{(2)}_{2l-1}(l\geq3):\, \alpha_i=\epsilon_i-\epsilon_{i+1},\quad i=1,2,\ldots, l-1,\quad
\alpha_l=2\epsilon_l,\e$$
$$ D^{(2)}_{l+1}(l\geq 2):\, \alpha_i=\sqrt2 (\epsilon_i-\epsilon_{i+1}),\quad i=1,2,\ldots, l-1,\quad
\alpha_l=\sqrt2 \epsilon_l,\e$$
$$E_6^{(2)}:\, \alpha_i=\sqrt 2 \{\half(\epsilon_1-\epsilon_2-\epsilon_3-\epsilon_4),\epsilon_4,\epsilon_3-\epsilon_4,\epsilon_2-\epsilon_3\},\e$$
$$D_4^{(3)}: \alpha_1=\epsilon_1-\epsilon_2,\quad \alpha_2=-2\epsilon_1+\epsilon_2+\epsilon_3,\e$$
where $\epsilon_s$, $s=1,2,\ldots$ are orthogonal unit vectors. 

We define the array,
$$K_{i,j}^{a,b}=\half \alpha_a\cdot\alpha_b [\min(i,j)-i j/m],\e$$
where $m$ is the level. We also use $a_0$ which is equal to one except for the algebra $A_{2l}^{(2)}$,
$l=1,2,\ldots$, which is set equal to two. We define
$$t_a=\max(\alpha_a^2,a_0).\e$$

Denote by $M$ the root lattice spanned over the integers by $\alpha_a,\, a=1,2,\ldots,l$. We define
the linear map
$${\cal I} (\alpha_a)=\delta_a \alpha_a,\e$$
where $\delta_a$ is equal to one except for the algebra $A_{2l}^{(2)}$ and $a=l$ (the long root)
where it is equal to two. We use the Pochhammer symbol,
$$(q)_n=\prod_{j=1}^n (1-q^j),\e$$
where $n$ is an integer. Let $q_a=q^{t_a}$. We use the set of integers,
$n_j^{(a)}$, where $a=1,2,\ldots,l$ and $j=1,2,\ldots,m-1$.
We set
$$(q)_{\bf n}=\prod_{a=1}^l \prod_{j=1}^{m-1} (q_a)_{n_j^{(a)}}.\e$$

The string functions of the twisted affine algebras are defined by the sum \r\Kac,
$$c_\lambda^\Lambda(\tau)=q^d \sum_{n=0}^\infty p_n e^{2\pi i \tau n},\e$$
where $d$ is some dimension, $q=\exp(2\pi i\tau)$ and $p_n$ is the multiplicity of a  weight
at grade $n$
and at the highest weight representation $\Lambda$ and the finite weight $\lambda$.
Here $i=\sqrt{-1}$ and the level is denoted by $m$.

We can state now our main result which is a Rogers--Ramanujan type expression for the string 
functions. We define the sum,
$$N_{m,\lambda}^\Lambda(q)= \prod_{a=1}^l (q_a)_{\infty}^{-1} \sum_{n_j^{(a)}\geq 0}
{q^{{\cal L}({\bf n})+{\cal V}({\bf n}) } \over (q)_{\bf n}},\e$$
where
$${\cal L}({\bf n})=\sum_{a,b=1}^l \sum_{i,j=1}^{m-1}  n_i^{(a)} n_j^{(b)} K_{i,j}^{a,b}.\e$$
The sum is limited to the integers $n_j^{(a)}$ which obey,
$$\sum_{a=1}^l \sum_{j=1}^{m-1} j\alpha_a n_j^{(a)}={\cal I} (\lambda-\Lambda) \mod m M,\e$$
where the linear map $\cal I$ was defined in eq. (10).

The string functions are given by,
$$c_\lambda^\Lambda(q)=N_{m,\lambda}^\Lambda(q^{1/a_0}),\e$$
For the $\Lambda$ and ${\cal V}({\bf n})$ which will be specified below  in three cases.

Case 1: Here we take $\Lambda=0$, the singlet representation and ${\cal V}({\bf n})=0$, for all the
twisted algebras. This 
is the expression of Hatayama et al. \r\Hata.

Case 2:  We need the matrix which is the inverse Cartan matrix of $SU(m)$, where $m$ as before
is the level,
$$B_{i,j}=\min(i,j)-ij/m,\e$$
Here, $i,j=0,1,\ldots ,m-1$.
We have an expression for the string functions for all the short weights which have a mark one.
For the algebra $A_{2 l-1}^{(2)}$ this is the weight $\Lambda=r\Lambda_1$, where $\Lambda_1$ is the first fundamental weight. We set in this case $a=1$. The other case is the algebra $D_{l+1}^{(2)}$.
Here we take $\Lambda=r\Lambda_l$. Here we set $a=l$. These two algebras are the only ones 
which have such roots.
For these weights we have the GRR expression, eq. (14), where we choose,
$${\cal V}({\bf n})=-\sum_{i=1}^{m-1} B_{m-r,i} n_i^{(a)}.\e$$ 

Case 3: We consider the algebra $A_{2 l}^{(2)}$, where $l=1,2,\ldots$. We take for the weight,
$\Lambda=r\Lambda_l$, which corresponds to the long root of the algebra, and
$r=1,2,\ldots,m/2$. We define the vector,
$$F_i=\min(i,r)-i/2+s_i,\e$$
where 
$s_i$ is $\half$ for odd $i$ and is zero otherwise. Here we take
$i=1,2,\ldots 2 r-1 $ and we take $F_i=0$ otherwise. We define the vector,
$$M_i=F_i-B_{r,i},\e$$
and we take
$${\cal V}({\bf n})=2\sum_{i=1}^{m-1} M_{m-i} n_i^{(l)}.\e$$

This completes our description of the GRR for the twisted affine algebras.
The expression holds for all the $\lambda$, in eq. (14) and the $\Lambda$ specified above.
For the algebra $A_2^{(2)}$ (case 3), eq. (14) gives all the string functions.

To verify our conjectures (cases 2 and 3), we use Freudenthal-- Kac formula
\r\Kac, which we computerized in the fortran program ALGEBRA (written in a collaboration with
A. Abouelsaud).

Let us give an example of $A_2^{(2)}$ with $\Lambda=\Lambda_1$ at level $m=3$ (case 3).
Here the matrix $K$ is
$$K={1\over 3} \pmatrix{ 2 & 1\cr 1 & 2 \cr}.\e$$
Then, our conjecture for the string functions becomes, for $\lambda=\Lambda$,
$$(q)_{\infty} \chi^\Lambda_\Lambda(q)=\sum_{m_1,m_2\geq 0\atop m_1+2 m_2=0\mod 3}
{q^ {\{ m_1,m_2\}.K.\{m_1,m_2\} /2 + (m_1-m_2)/3 } \over
(q)_{m_1} (q)_{m_2} },\e$$
Using a Mathematica program we can compute the expression, eq. (24), up to order 8 and we find
$$\chi_\Lambda^\Lambda (q)=1+2 q +6 q^2 +12 q^3 + 26 q^4+48 q^5+90 q^6+156 q^7 +270 q^8+\ldots.\e$$
From the program ALGEBRA we find the exact same expression up to grade 8.

Let us give an example of the algebra $D_3^{(2)}$ at level two with $\Lambda=\Lambda_2$
(case 2), and $\lambda=\Lambda$.
The matrix of scalar products is in this example,
$$K=\pmatrix{4 & -2\cr -2 & 2 \cr}.\e$$
From the general conjecture, eq. (14), we find for the character
$$(q)_{\infty} (q^2)_\infty  \chi^\Lambda_\Lambda(q)=\sum_{m_1,m_2\geq 0\atop m_1,m_2=0\mod 2}
{q^ {\{ m_1,m_2\}.K.\{m_1,m_2\} /4 - m_2/2 } \over
(q^2)_{m_1} (q)_{m_2} },\e$$
Using Mathematica, this expression is evaluated to
$$\chi_\Lambda^\Lambda(q)=1+3 q+8 q^2+19 q^3+41 q^4+83 q^5+161 q^6+ 299 q^7+\ldots.\e$$
Using the program ALGEBRA we find the exact same expression for the string function,
up to grade $7$.

\ack
I thank A. Abouelsaud for his collaboration in the program ALGEBRA and J. Ramos for remarks on the
manuscript.

\refout

\bye